\documentstyle[11pt,fleqn]{article}
\date{May 1996}

\begin{document}
\title{\LARGE \bf Pregeometric modelling of the spacetime phenomenology}  
\author{{Reginald T. Cahill  and     Christopher M. Klinger}\\
  {Department of Physics, Flinders University
\thanks{E-mail: Reg.Cahill@flinders.edu.au, Chris.Klinger@flinders.edu.au}}\\ { GPO Box
2100, Adelaide 5001, Australia }}

\maketitle

\begin{center}
\begin{minipage}{120mm}
\vskip 0.6in
\begin{center}{\bf Abstract}\end{center} {At present we have only the very successful but
phenomenological Einstein geometrical modelling of the spacetime phenomenon. This
geometrical model provides a `container'  for other theories, in particular the quantum field
theories. Here we report progress in developing a {\em Heraclitean Quantum System}. This is a
particular  pregeometric theory  for  space and time  in which no classical
or geometric structures are assumed, but rather the emergence of such phenomena is
sought.   }

\end{minipage} \end{center}

\vspace{50mm}
\hspace{10mm} {\em Ta panta rei - all is flux}
\vspace{3mm}

\hspace{10mm}  Heraclitus of Ephesus, sixth century BC

\newpage

\noindent  {\bf 1. Introduction}  \hspace{5mm}
 At present we have no theory of the phenomena of time and space.  Rather we have a very
successful  phenomenology given to us by Einstein. We regard Einstein's model as a
phenomenology for the simple reason that in setting up this model one makes very explicit
assumptions about time and space. For example
one of the key Einstein assumptions was to assume, in addition to the very phenomenon of
time,  that time is  local, which contrasted sharply with Newton's assumption of
a global time.   We take the defining   indicator  of a theory to be the property
that it predicts  phenomena, but does not have those phenomena explicitly or even covertly
built  into its axioms. Well known examples of successful theories include the atomic
theory: it predicted the existence and properties of atoms, molecules, etc, but  contained
only electrons and nuclei in the axioms. A second example is that of nuclear physics: Quantum
Chromdynamics (QCD) begins with quarks and gluons  and predicts the  phenomena of hadrons,
nuclei etc. In this example there was a long period of phenomenological modelling in
which hadrons were described by effective actions, or equivalently Hamiltonians,  involving
hadronic fields. The structure of these effective actions was obtained by appealing to
various symmetries that appeared to be manifested in the hadronic data. These examples
illustrate the idea of emergent phenomena.

A feature of the Einstein model is that it is a geometrical model,  and is a generalisation
of the geometrical modelling  by Galileo and Newton. These models  build upon the
different Ancient Greek models of Pythagoras, Parmenides and Democritus. This modelling is so effective and
persuasive that there is a tendency to confuse the phenomena  of time and space with the
geometrical modelling. For example the modelling of time, whether local or global, by the
real number line is often implicitly assumed to be an actual property of the phenomenon
of time. 

The present standard model of physics, while successful, has a very strange three stage 
structure. First one constructs a classical geometrical spacetime structure. Second,
various classical fields are attached to this geometrical structure, and finally, in the
third stage one {\em quantises} the matter fields.    As an afterthought, one might even
attempt to derive the classical behaviour of large quantum systems by means of some
classicalisation argument. What one sees in this structuring is  an incomplete separation of
the historical development of the subject from a proper theoretical structure. In a mature
theory one would expect to see the classical features as emergent properties of some
abstract quantum system which itself does not contain classical structures. We call such
systems Heraclitean Quantum Systems (HQS) after Heraclitus of Ephesus (540-480 BC) who
appears to have anticipated such ideas by some 2500 years. He argued that  common sense is
mistaken in thinking that the world consists of stable things; rather the world is in a
state of flux. The appearance of `things' depends upon this flux for their continuity and
identity.  What needs to be explained,
Heraclitus argued, is not change, but the appearance of stability. We suggest that the
success of the standard model in its present three stage form is a clear indication of some
extremely robust mean-field type phenomena  arising within a HQS. Rather than a co-evolving
unified system, our three-stage modelling amounts to first evolving space with at most
only a macroscopic input from the  matter energy density, and then in a second
computational sweep we evolve the quantum matter. While  certainly a very effective
approximation it nevertheless makes it impossible for correlations in the quantum processing
of the spatial phenomenon to influence  the quantum matter processing.  Indeed, while  it is
this approximation which makes possible local field theories, it also renders these same
theories with short distance problems.  

In recent years much effort has been put into attempts to quantise gravity. To us this
seems a suspect procedure for discovering deeper theories.  Quantisation is not a
fundamental physical process, rather it is a guessing procedure that has been invoked
somewhat fruitfully in the last 70 years. To give a recent counter example we note that the
quark-gluon quantum system was not obtained by {\em quantising} the classical hadronic 
phenomenological field theory of the 1960s. Rather the next level down from hadrons was
obtained by some inspired guessing. The  problems that then arose were the  
demonstration that hadronic laws, in the form of an effective action description,  could be 
extracted from the quark-gluon system, and also the experimental study  of hadronic systems
to reveal signatures of the quark-gluon subsystem.

Of considerable current interest is the process of classicalisation. In this one attempts
to deduce a special emergent behaviour of large quantum systems. In such systems classical
behaviour is a weird  and poorly understood phenomenon. The obvious fact that we 
modelled classical behaviour first does not deny the fact that it is a secondary effect.  
Basic quantum mechanics is still bedevilled by the metaphysical fix-ups that were invoked
in the early days of quantum theory  when studying the transition from a simple to a complex
large scale quantum system that occurs during the measurement process. Fortunately 
classicality is now seen as a physical process  requiring detailed dynamical analysis.
Quantisation  is not de-classicalisation. By quantising some classical
emergent phenomenon we do not recover the deeper quantum system that produced
that classicality.

The need to construct  a non-geometric theory to explain the time and space
phenomena has been strongly argued by Wheeler \cite{Wheeler80}, under the name of
pregeometry. Gibbs
\cite{Gibbs} has recently compiled a literature survey of such attempts.  They include:
   cellular automata,
lattice field theories,
 quantum metric spaces,
 causal nets,
 poset models,
 simplicial quantum gravity,
 fractals,
 topological quantum field theory,
 field theory on a complex cell,
 spin networks,
 twistor theory,
 signal spaces,
 non-commutative geometry and
 event-symmetric space-time.
 Isham \cite{Isham} has recently discussed the possibility that spacetime is indeed a
phenomenological construct, and not fundamental.  

Here we give a brief outline of some of the insights that have been obtained for a
particular Heraclitean Quantum System, which is a  pregeometric type model with no
classical structures assumed in its axioms. In many pregeometric models  residual
classical and phenomenological structures are retained. Our  HQS is an abstract
Grassmannian algebraic system, which is based upon some of the insights gained from the
derivation \cite{RTC} of the emergent hadronic  phenomena for the quark-gluon system. The
nature of this analysis is briefly discussed in section 2, and highlights the idea of
action sequencing induced by dynamically determined  changes of
functional integration variables. In section 3 a particular HQS is presented together with
some insight into how  spacetime  might arise.

\vspace{5mm}

\noindent {\bf 2  Action Sequencing}  \hspace{5mm} QCD provides us with a fine example of
the emergence of complex  effective theories. Some parts exhibit  an induced geometrical
form, while overall we see the idea of action sequencing that is an integral  part of the
derivation of emergent phenomena, and also the importance of condensate effects. These
results have been achieved  using the Functional Integral Calculus (FIC) 
\cite{RTC} which most powerfully takes advantage of the functional integral formulation of
quantum field theories.

   QCD essentially involves the functional integral in  (\ref{eq:2.1}) for the
vacuum persistence amplitude  in the presence of sources $J$ (which are not shown on the
RHS). At low energies or long wavelengths we only observe hadronic degrees of freedom, and
not the quark and gluon fields.  In this respect we expect QCD to be  archetypal: in HQS we
do not expect to observe the fundamental defining  algebraic elements. Their usefulness will
rest solely upon their role in successfully predicting a large amount of higher level
observable phenomena. The derivation of the low energy form of QCD, namely the hadronic
form, is outlined in (\ref{eq:2.1})-(\ref{eq:FIC}). As expected  this derivation is
not exact. A useful step is that of approximating QCD by the Global Colour Model
(GCM) \cite{RTC}.

\begin{equation}  <0 \mid 0>_J=\int {\cal D}\overline{q}{\cal D}q{\cal
D}Aexp(-S_{QCD}[A,\overline{q},q])
\label{eq:2.1}\end{equation}

\begin{equation}\mbox{\ \ }\hspace{14mm}\approx\int {\cal D}\overline{q}{\cal D}q{\cal
D}Aexp(-S_{GCM}[A,\overline{q},q])   \mbox{\ \ \ \ (Global Colour Model)}
\label{GCM}\end{equation}
  
\begin{equation}
\mbox{\ \ }\hspace{14mm} =\int {\cal D}{\cal B}{\cal D}{\cal D}{\cal D}{\cal D}^{\star}exp(-
S_{bl}[{\cal B}, {\cal D}, {\cal D}^{\star}]) \mbox{\ \ \ \ (bilocal fields)}
\label{eq:2.7}\end{equation} 

\begin{equation} \mbox{\ \ }\hspace{14mm} =\int{\cal D}\pi...{\cal D}\overline{N}{\cal
   D}N...exp(-S_{had}[\pi,...,\overline{N},N,..]) 
  \mbox{\ \ \ \ (local fields) }\label{eq:FIC}\end{equation} 

The derived hadronic 
action that finally emerges from this action sequencing, to low order in fields and
derivatives, has the form 
\begin{eqnarray*} \lefteqn{ S_{had}[\pi,...,\overline{N},N,..] = }\\&
 &\hspace{25mm}\int d^4x
tr\{\overline{N}(\gamma.\partial+m_N+  \Delta m_N-  m_N\surd 2i\gamma_5\pi^a{\cal
T}^a+..)N\}+\\ & &
\hspace{5mm}+\int d^4x\left( \frac{f_{\pi}^2}{4}tr(\partial_{\mu}U\partial_{\mu}U^
{\dagger})+\kappa_1tr(\partial^2U\partial^2U^{\dagger})+
\frac{\rho}{2}tr([{\bf 1} -\frac{U+U^{\dagger}}{2}]{\cal M})+\right.\\ & &   
\hspace{25mm}
\left. \hspace{15mm}+\kappa_2tr([\partial_{\mu}U\partial_{\mu}U^{\dagger}]^2)
+\kappa_3tr(\partial_{\mu}U\partial_{\nu}U^{\dagger}\partial_{\mu}U \partial_{\nu}U^
{\dagger})+\right.
\\ & &\left.\hspace{5mm}+
\frac{f_{\omega}^2}{2}[ -
\omega_{\mu}\Box\omega_{\mu} +(\partial_{\mu}\omega_{\mu})^2
+m_{\omega}^2\omega_{\mu}^2] +
\frac{f_{\rho}^2}{2}[ -
\rho_{\mu}\Box\rho_{\mu} +(\partial_{\mu}\rho_{\mu})^2
+m_{\rho}^2\rho_{\mu}^2]+\right.\\    & &\left.\hspace{31mm}-f_{\rho}f_{\pi}^2g_
{\rho\pi\pi}\rho_{\mu}.\pi \times\partial _{\mu}\pi
-if_{\omega}f_{\pi}^3\epsilon_{\mu\nu\sigma\tau}\omega_{\mu}\partial_{\nu}
 \pi . \partial_{\sigma}\pi\times \partial_{\tau}\pi+\right.\\ & &\left.\hspace{65mm}
-if_{\omega}f_{\rho}f_{\pi}
G_{\omega\rho\pi}\epsilon_{\mu\nu\sigma\tau}\omega_{\mu}\partial_{\nu}
\rho_{\sigma}.\partial_{\tau}\pi+\right.\end{eqnarray*}
\begin{equation}\left.\hspace{40mm}+
\frac{i}{80\pi^2}\epsilon_{\mu\nu\sigma\tau}tr(\pi.F\partial_{\mu}
\pi.F\partial_{\nu}\pi.F\partial_{\sigma}\pi.F\partial_{\tau}\pi. F)
+......\label{eq:hadact}\right)\end{equation}

This shows that the emergent hadronic phenomena are very rich and complex; that is why
the nucleus is so much more complicated than atoms.

We see in the above the powerful notion of {\em action sequencing}  
\begin{equation}S_{QCD}[A,\overline{q},q]\rightarrow S_{GCM}[A,\overline{q},q]
\rightarrow S_{bl}[{\cal B}, {\cal D}, {\cal
D}^{\star}]\rightarrow S_{had}[\pi,...,\overline{N},N,..]
\label{eq:actseq}\end{equation}
Each change of functional integration field  variables, and these are mandated  by the
dynamics, generates a new effective action for those field variables.  It is only the final
hadronic variables and their induced effective action that   allows us to relate  QCD
to the experimental data. Even the hadronic form in  (\ref{eq:FIC})  requires further
evaluation to produce the physical hadrons, since (\ref{eq:FIC})  involves the so-called
core or constituent states. The final hadronic functional integration dresses each of these
core states with a cloud of other hadrons, mainly low mass mesons.

A key intermediate step is the determination of the minimum of the action in (\ref{eq:2.7}) 
\begin{equation}
\frac{\delta S_{bl}[{\cal B}, {\cal D}, {\cal
D}^{\star}]}{\delta {\cal B}}=0,...\label{eq:QCDvac}\end{equation}
which  has a solution with  ${\cal B}\neq 0$ and  gives the $\overline{q}q$
{\em condensate}  effect. This simply means that the induced effective action has a
non-trivial minimum away from the perturbative ${\cal B}= 0$ point.  Similar effects occur
in superconductivity.  This condensate effect is one of the most important
dynamical effects in QCD and goes a long way in explaining the nature of hadrons. In
particular it generates a running mass for the constituent quarks, and leads to the
constituent quark mass of some $300MeV$. A recent account is given in
\cite{CG}. The  structure of the condensate and the consequent structure of the hadrons is
determined by the   gluon correlations. At the end of the  calculation of the derivative
expansion  we, in effect,  suppress any explicit mention of the internal structure of the
hadrons, resulting in  local couplings of local fields -  the emergent hadronic phenomenon.
 
 Hadrons may be viewed as  deviations in the structure of  the  condensate. The lowest
mass hadrons correspond to those  deviations in the flattest directions of  the
effective action for the bilocal fields, in (\ref{eq:2.7}).  These correspond to the pions.
If the quark current masses are zero then these are directions in which the action is
strictly flat, and the resulting massless mesons are known as Nambu - Goldstone  (NG)
bosons. These massless modes are represented in (\ref{eq:hadact}) by the matrix
$U(x)=exp(i\surd 2\pi^a(x)F^a)$ where  the $\{F^a \}$ are the generators of the $SU(N_f)$
flavour symmetry group. The NG boson fields $\pi(x)$ form homogeneous Riemann coordinates
for this vacuum manifold, which has the form of a coset space. The internal
structure of the pions is intimately related to the structure of the condensate. Thus the  
long range part of the nuclear force is determined by the  near degeneracy of the
condensate  equations (\ref{eq:QCDvac}). See \cite{Frank} for a recent analysis of the
pion sector of the GCM. We end this section by giving an insight into the nature of the 
condensate deviations. If 
$\overline{\cal B}_0(x,y)$  is a particular solution of (\ref{eq:QCDvac}),  possibly having
degenerate solutions, then the idea of a condensate  deviation is given by
\begin{equation}
{\cal B}(x,y)=\overline{\cal B}_0(x,y)+\sum_a \phi_a(\frac{x+y}{2})\Gamma^a(x-y)
\label{eq:dev}\end{equation}
in which we expand the $x-y$ dependence of ${\cal B}(x,y)$ into a complete set
 $\Gamma^a(w)$, with the  $\phi_a(z)$ as
expansion coefficients. We may change the  variables of integration  from the ${\cal B}(x,y)$
to the $\phi_a$. In the GCM the $\Gamma^a(w)$ are chosen in order to  diagonalize the 2nd
order terms arising  when the bilocal effective in (\ref{eq:2.7}) is expanded about
$\overline{\cal B}_0(x,y)$. This essentially leads to the
$\Gamma^a(w)$ being solutions of Bethe-Salpeter equations, and describing the internal
structure of $\overline{q}q$ core states. The fields
$\phi_a(z)$  describe the `centre-of-mass' motion of these mesonic bound states. Fields of
this type occur in  (\ref{eq:FIC}).  This expansion
procedure leads to  a bosonisation of QCD. A  detailed formal derivation and generalization
to introduce diquarks and baryons is given in \cite{RTC}, leading to the hadronisation of
QCD.  Recent results and discussion are given in \cite{CG}.

\vspace{3mm}

 \noindent {\bf 3.  Heraclitean Quantum Systems} \hspace{5mm}  A HQS has no classical
structures or concepts built into the axioms. We will consider a  model with only abstract
algebraic elements. This algebra  is taken to be a Grassmann
algebraic system. Such algebras are used very effectively to model the fermionic sector of
the standard model, though in these applications they are mutli-component  local
`algebraic fields' attached to some spacetime manifold. Here we ask whether such an
abstract system can induce our very successful  spacetime phenomenology together with the
quantum matter modelling of our present standard model.  Can we adapt the bosonisation
techniques developed  in the GCM to  discover any classical features that such a system might
possess? The Grassmann algebra  retains, as an assumed  intrinsic property, an abstract form
of the Pauli Exclusion Principle, which is realised via the anti-commuting property of
the algebra. The algebra is a set of
$2N$ elements, with  eventually $N \rightarrow \infty$.
\begin{equation}\{M_i,i=1,2,...,2N 
\}=\{\overline{m}_i,m_i,i=1,2,...,N\}\label{eq:xx1}\end{equation}
 and, by definition, mutually anticommuting
\begin{equation}M_iM_j=-M_jM_i \mbox{\ \ \  so that \ \ \ }
M_iM_i=0\label{eq:xx2}\end{equation}  
The subscript explicitly labels these simple elements: there are no hidden  indices. The
distinction between
$\overline{m}_i$ and
$m_i$ only arises when some form for the `action' $S_{HQS}[\overline{m},m]$ is specified.
We name $\overline{m},m$ monads after  Leibniz \cite{Leibniz}.
An  abstract notion of  `correlation' is defined by 
\begin{equation}G^{i,..}_{j...}= {\cal
G}[\overline{m}_im_j...e^{-S_{HQS}[\overline{m},m]}]\label{eq:xx3}\end{equation} which
involves Grassmann `integration'.
${\cal G}[...]$ is a purely algebraic process \cite{Berezin}. 
 Such extremely primitive `correlations'
are not expected to have in themselves any phenomenological significance. The phenomena we
are looking for must be manifested with respect to the many complex co-phenomena that a HQS
must produce in order to be a viable theory. That is, we must  develope an `internal view' of
the emergent phenomena. Unlike our present quantum modelling we cannot take an `external
view'.  The basic `processing' of the HQS is based on the assumption that the Grassmannian
modelling of the fermionic sector of the standard model, with its quantum correlations being
determined by a Grassmannian `integration',  is a residual property of an underlying  HQS. 
Indeed we have seen in the GCM that this  Grassmannian  modelling
of the quarks re-emerges, at a higher level, as a Grassmannian description of the baryons.

A more general `correlation' is 
\begin{equation}G[F]={\cal
G}[F[\overline{m},m]e^{-S_{HQS}[\overline{m},m]}]\label{eq:xx4}\end{equation} where $F$ is
some function of all the elements. To define the integration process we expand
the `integrand' as a polynomial
\begin{equation}F[\overline{m},m]e^{-S_{HQS}[\overline{m},m]}=1+\sum
c_im_i+...+c_L\overline{m}_1\overline{m}_2....\overline{m}_Nm_1m_2...m_N\label{eq:xx5}\end{equation}  
The sum of the terms of highest order  has been 
written  in some standard order.
 Then by definition,  
\begin{equation}{\cal
G}[F[\overline{m},m]e^{-S_{HQS}[\overline{m},m]}]=c_L\label{eq:xx6}\end{equation}

A particular Grassmann integration that can be explicitly performed is \cite{Berezin}
\begin{equation}
{\cal G}[e^{-\sum M_iA_{ij}M_j}]=\mbox{Pf}(2A)
\label{eq:special}\end{equation}
where $A$ is an antisymmetric matrix, and where the Pfaffian Pf$(A)$ is the square root of
the determinant of an antisymmetric matrix $A$, in the sense that Pf$(A)^2=\mbox{det} A$.

 Extending the Grassmann algebra to include `sources'
$\overline{l}_i,l_i$, a  generating functional is introduced
\begin{equation}Z[\overline{l},l]= {\cal
G}[e^{-S_{HQS}[\overline{m},m]-\overline{l}m-\overline{m}l}]\label{eq:xx7}\end{equation}

Let us consider the particular HQS defined by the   quartic action  
\begin{equation}
S_{HQS}[\overline{m},m]=-
\frac{1}{2}\overline{m}.m\overline{m}.m=\sum_{i>j}\overline{m}_im_j\overline{m}_jm_i
\label{eq:xx9}\end{equation}
No notion of locality is permissible, so all elements are in `interaction'. The apparent
dominance of local interactions must be emergent. 
 The action has a large  invariance group: $m\rightarrow Um, \overline{m}\rightarrow
\overline{m}U^{-1}\label{eq:Usym}$.

 Consider a  bosonisation along the line of the GCM bosonisation in QCD.
We can put $Z$ in the form
\begin{equation}Z[\overline{l},l]= {\cal
G}[\int{\cal
D}Be^{-S_{Bmm}[B,\overline{m},m]-\overline{l}m-\overline{m}l}]\label{eq:BB10}\end{equation}
where \begin{equation}
S_{Bmm}[B,\overline{m},m]=\frac{1}{2}\sum_{i,j}B_{ij}B_{ij}-
\sum_{i,j}B_{ij}(\overline{m}_im_j-\overline{m}_jm_i)\end{equation}
with $B_{ij}=-B_{ji}$ and real. This is easily checked on doing the gaussian
$B$-integrations. We may now explicitly perform the ${\cal G}$ process giving

\begin{equation}Z[\overline{l},l]=\int{\cal
D}Be^{-\sum_{i>j}B^2_{ij}+\mbox{\small{TrLn}}(B)+\overline{l}B^{-1}l}\label{eq:BB11}\end{equation}
  
 The Grassmann algebraic aspects
are now contained in $\mbox{TrLn}$ and the $\overline{l},l $ algebra.  The algebraic ${\cal
G}$ process has now been given a representation involving the sum over all possible
$B$ configurations.  This  is the Heraclitean `flux', and is an axiomatic aspect of the
HQS. What is sought is the emergence of stability and `things'. Or as Wheeler calls it
``it from bit''. 

 The induced action is 

\begin{equation}
S_{C}[B]= \sum_{i>j}B^2_{ij}-\mbox{TrLn}(B)
\end{equation}
  Here `$B$' is the analogue of bilocal fields in QCD. The advantage of the bosonisation is
that it is more amenable to our well honed  analytical techniques. As well our
present day modelling has indicated the extraordinary success of mean field or
smoothing approximations. We expect these to be accessible via the bosonisation. 
However note that the bosonisation does not preclude the emergence of complex fermionic
components, as again illustrated by the GCM. The bosonisation technique now proceedes with
an analysis of the  minimum  of the induced action. This  identifies the most
significant part of the
$B$ integrations. 
\begin{equation}\delta S_{C}[B]/\delta B=0\label{eq:vaceqn}\end{equation}
gives $B=-B^{-1}$ - the `condensate' equation, with solutions $\overline{B}$,   analogous to
the gap equation in superconductivity, and to the  condensate equation in QCD. 
The general solution is
$\overline{B}=R\overline{B}_0R^{-1}$ where $R$ is an arbitrary real  orthogonal matrix and
$\overline{B}_0$ is the block diagonal matrix
\begin{equation}\overline{B}_0=\left(\begin{array}{rrrrr}
0 & +1 & 0 & 0 & ...\\
-1 & 0 & 0 & 0 & ...\\
0 & 0 & +1 & 0 & ...\\
0 & -1 & 0 & 0 & ...\\
            \\
\end{array}\right)\label{eq:array}\end{equation}
 Hence the condensate is highly degenerate.  The $R$ transformation 
`switches'  monad pairings.  The degeneracy of the condensate dominates the
$B$-fluctuations.   We now search for signs of an  emergent   spacetime phenomenon, quantum
fields, etc.

 Consider the  `{\em nihilo}  $\rightarrow$ {\em nihilo} amplitude'
   \begin{equation}
<{\cal N}\mid{\cal N}>=Z[0]=\int {\cal
D}Be^{-S_{C}[B]}\label{eq:BB12}\end{equation}
 and the deviation from  $\overline{B}_0$, in which the $\Gamma^a$ must be a complete set.
\begin{equation}
B_{ij}=\overline{B}_{0ij}+\sum_a\phi_a\Gamma^a_{ij}
\label{eq:deviate}\end{equation}
Our first choice for the  $\Gamma^a_{ij}$ is the following:
 Set `$a$' to be a serial index $a \equiv (IJ)$, and with $\Gamma^a_{ij}=-\Gamma^a_{ji}=+1$
if $I=i$ and $J=j$, otherwise $\Gamma^a_{ij}=0$. These $\Gamma^a_{ij}$ form a complete set
for the expansion, with expansion coefficients $\phi_a$.  Then, changing variables of
integration (the Jacobian is a constant and can be ignored),
\begin{equation} <{\cal N}\mid{\cal N}>=\int {\cal
D}\phi e^{-S_{C}[\overline{B}_0+\phi.\Gamma]}\label{eq:phieq}\end{equation}
This choice is essentially equivalent to the defining $B_{ij}$ integrations. 
As usual with degenerate condensates we make the superselection assumption that we can work
in the neighbourhood of one condensate point, say $\overline{B}_0$, and  expand $S_C$ in
powers of $\phi_a$
\begin{equation}
S_C[\overline{B}_0+\phi.\Gamma]=S_C[0]+\sum_{ab}\phi_a\phi_bK_{ab}+
\sum_{abc}\phi_a\phi_b\phi_cK_{abc}+..
\label{eq:expan}\end{equation}
where there is no linear term because of the condensate equation (\ref{eq:vaceqn}).
These variables of integration affect only small numbers of  $\overline{m}-m$ pairings.
They are too primitive to be able to reveal any complex emergent behaviour. Nevertheless we
can partly analyse them by  choosing new variables of integration by diagonalising the
quadratic term in (\ref{eq:expan}), giving
\begin{equation}
<{\cal N}\mid{\cal N}>=\int {\cal D}\Phi e^{-S_C[0]-\sum_{a}\Phi_a\Phi_a\lambda_a-
\sum_{abc}\Phi_a\Phi_b\Phi_cK'_{abc}+...}
\label{eq:Phi}\end{equation}
This change of variables is equivalent to a new choice for the  $\Gamma^a_{ij}$.
Approximately one half of the eigenvalues $\lambda_a$ have value zero:  these correspond to
the `massless' NG modes, i.e. deviations in the tangent plane to the condensate manifold.
The remaining $\lambda_a$ are all non-zero and equal: these `massive' modes correspond to
deviations perpendicular to the condensate manifold.
In QCD the analogue of the $\Phi_a$ modes are $\overline{q}q$  meson core-state modes, and
the diagonalisation procedure is there the Bethe-Salpeter equation. Because of the
peculiarities of QCD the hadrons contain either two constituent quarks (mesons) or three
constituent quarks (baryons) together with secondary mesonic dressings of these core states. 
However in this HQS we are interested in multi-monad modes, within which we hope to find
evidence of classical structures.  For this purpose the above two possible choices of
integration variables are not  helpful. 

We now consider yet a third choice of integration
variables. In (\ref{eq:deviate}) consider a new set of $\Gamma^a$: $\Gamma^a_{ij}= +1 $ with
probability $
\frac{p}{2}$ or 
$ -1 $ with probability $\frac{p}{2}$, and  $ =0 $ otherwise, i.e. with
probability $q=1-p$. In some sense each such  $\Gamma^a$ corresponds to some
random multi-monad excitation of the condensate. We need this set to be complete. With the
extreme choice $p=0$  only one trivial $\Gamma^a$ is formed.  Similarly, if $p=1$ we form
only $\Gamma$'s with all off-diagonal entries being $+1$ or $-1$.  However if
$p<<1$ then the
$\Gamma^a$ have sparse non-zero entries, and approximate a complete set. Hence changing
to these variables in (\ref{eq:phieq}), and using $G=\{G_a\}$ as the new variables of
integration,
\begin{equation} <{\cal N}\mid{\cal N}>=\int {\cal D}G 
e^{-S_{C}[\overline{B}_0+G.\Gamma]}
\label{eq:Geq}\end{equation}

This new set of multi-monad $\Gamma^a$ has a very  interesting interpretation.
To each such $\Gamma^a_{ij}$ matrix we can associate a random graph: consider the indices
$i$ or $j$ as labelling the `points' or `nodes' of a graph, in which two points $i$ and $j$
are linked if $|\Gamma^a_{ij}|=1$. Such a graph is in general composed of disconnected
pieces.  Although we have no a priori background geometry we can nevertheless define one
measure of distance between points within a connected piece by counting the minimum number
of links connecting the points. Nagels \cite{Nagels} has considered the probability
distribution of such distances in  connected random graphs. Let $D_k=1,2,3,..$  be
the number of points a distance $k=0,1,2,3,..$ from a particular arbitrary point,
called the origin. So $D_0=1$ (by definition), $D_1$ is the number of adjacent points, etc. 
For $p <<1$ the shape of a connected random graph, as defined by the (relative) probability
distribution of distances,  is given by
\begin{equation}
P[D_k]=\prod^L_{i=1}\frac{(D_{i-1})^{D_i}}{D_i!}
\label{eq:shape}\end{equation}  
where $L$ is the maximum distance of any point from the origin. Further
\begin{equation}
\sum_{k=0}^LD_{k}=N_c
\label{eq:total}\end{equation}
where  $N_c$ is the number of points in the connected random graph.
The most probable distribution, i.e. the most probable connected graph, 
from maximizing $P[D_k]$ subject to the constraint (\ref{eq:total}) \cite{Nagels} is 
\begin{equation}
D_k\sim \frac{L^2\mbox{ln }L}{2\pi^2}\left[\mbox{sin}^2\left(\frac{\pi
k}{L}\right)-\frac{1}{3}\left(\frac{\pi}{L}\right)\mbox{sin}\left(\frac{2\pi
k}{L}\right).\mbox{ln sin}\left(\frac{\pi k}{L}\right) 
\right]
\label{eq:prob}\end{equation}
The remarkable property of the most probable distribution in (\ref{eq:prob}) is that the
resulting emergent structure closely resembles a three-dimensional closed space of positive
curvature, for we obtain from  (\ref{eq:total}) that
\begin{equation}
N_c \sim \left(\frac{\mbox{ln}L}{4\pi^2}\right)L^3
\label{eq:3D}\end{equation}
We also see the leading sin$^2$  term in (\ref{eq:prob}) characteristic   of the
hypersphere $S^3$. 

The partitioning of the random graphs into connected pieces is matched
by a corresponding partitioning of each $\Gamma^a$, so that for the integration variables
$G_a$ each subscript labels, in the most probable case,  some three-dimensional kind of
closed space, with internal structure specified by a $\Gamma^a$.  After expanding the
exponent in (\ref{eq:Geq}) in powers of $G_a$  and computing the trace summations,
 we are left with an induced effective action for the $G_a$
\begin{equation}
<{\cal N}\mid{\cal N}>=\int {\cal D}G  e^{-S[G]}
\label{eq:induced}\end{equation}
corresponding to a quantum `field' theory of interacting 3-spaces.
  This clearly has similarities  with some  programs that are currently
being pursued in quantising general relativity.  However it is not clear that our 3-spaces
are necessarily to be directly identified with the spatial section of a `universe', for one
might expect to see some further condensation processes for these `core' 3-space states
leading to a fractal or foamy spatial structure. From the point of view of quantising
gravity Wheeler first pointed out that quantum fluctuations in the metric would give
spacetime a foam-like structure. However in  attempted quantisations of  gravity  the
manifold  point of view is maintained with considerable mathematical difficulty. Our HQS is
not so constrained  and a whole range of possible short distance behaviour is possible.
Various short distance manifold descriptions have been studied by Hawking \cite{Hawking},
Coleman \cite{Coleman} and others.

 However the
outstanding problem is to show that the HQS can induce a time phenomenon: can we demonstrate
a natural classical sequencing - the basic phenomenon of time?  To this end we introduce a
complete set of functions $\{f_{\alpha}(G)\}$  for which
\begin{equation}\delta(G^2-G^1)=\sum_\alpha f_\alpha^*(G^2)f_\alpha(G^1)
\label{eq:compl}\end{equation}
 where now the superscript on $G^i$ labels different copies of
$\{G_a\}$.   Then
\begin{equation} <{\cal N}\mid{\cal N}>=
\int{\cal D}G^2\int{\cal D}G^1 e^{-\frac{1}{2}S[G^2]}\delta(G^1-G^2)e^{-\frac{1}{2}S[G^1]} 
\label{eq:xx14}\end{equation}
\begin{equation} \hspace{20mm}=\sum_\alpha\int{\cal D}
G^21.e^{-S^{(2)}[G^2]}f_\alpha^*(G^2)\int{\cal D}G^1f_\alpha(G^1) e^{-S^{(2)}[G^2]}.1 
\label{eq:xx15}\end{equation} where we define, in general,
$S^{(n)}[G]=\frac{1}{n}S[G]$.  Think of 
\begin{equation}
C_\alpha^{(2)}=\int{\cal D}G^1f_\alpha(G^1)e^{-S^{(2)}[G^1]}.1
\label{eq:xx16}\end{equation} 
as a transition amplitude, where  the `$1$'  represents the beginning of the `universe', i.e.
with all $G$ equally likely. Continue  inserting complete sets 
\begin{equation}
C_\alpha^{(2)}=\sum_\beta\int{\cal D}G^1f_\alpha(G^1)e^{-S^{(4)}[G^1]}f_\beta^*(G^1)
\int{\cal D}G^2 f_\beta(G^2)e^{-S^{(4)}[G^2]}.1
\label{eq:xx18}\end{equation}
\begin{equation}
C_\alpha^{(2)}=\sum_\beta A_{\alpha\beta}^{(4)}C_\beta^{(4)}
\label{eq:xx19}\end{equation}
\begin{equation} 
A_{\alpha\beta}^{(n)}=\int{\cal D}Gf_\alpha(G)e^{-S^{(n)}[G]}f_\beta^*(G) 
\label{eq:xx20}\end{equation}
More insertions give
\begin{equation}C_\alpha^{(2)}=\sum_{\beta\gamma...}A_{\alpha\beta}^{(n)}
A_{\beta\gamma}^{(n)}....C_.^{(n)}
\label{eq:xx21}\end{equation}
 which has the form of an all-inclusive quantum-like  multiple sequencing. But the time
phenomenon is about 
  restricted or classical sequencing, with  only some residual quantum phenomena.
 A possible macroscopic unique sequencing or history is a partition of \linebreak
$<{\cal N}\mid{\cal N}>$ such that different histories, by definition, have negligible
interference. They are decoherent. They are classical.  Hence we must look for a particular 
choice of complete set
$\{f_\alpha(G)\}$ for which some of the members generate decoherent and robust histories.
This has some resemblence to the consistent histories approach to standard quantum theory
by Griffiths \cite{Griffiths}, Omn\`{e}s \cite{Omnes} and Gell-Mann and Hartle \cite{GMH}.

There is a limit to the usefulness of these complete set insertions.
 For the  action $S^{(n)}$ becomes increasingly flatter, so that
fluctuations  or deviations from the condensate become  more 
extreme,  suggesting that  any time-like sequencing description has limited relevance
at very short time intervals.  This  resolves the objections of Parmenides and Zeno to 
infinite `information processing'. So in a HQS the modelling of time by the real number
line  will be limited by the nature of the fluctuation dominance that sets in at too fine a
resolution: the very concept of spacetime simply dissolves away into  a flux of
non-geometric fluctuations. At intermediate  scales the HQS would appear to produce a
spacetime modelling resembling a foamy or fractal structure.

 HQS is necessarily a  quantum cosmology. But  quantum cosmology is not just about the
beginning of the universe; it is about the ongoing evolution of the universe. 
It is a part of the classicalisation process that classical (continuum) differential
equations can be used to evolve 3-spaces etc, but that phenomenology is contingent upon the
underlying Heraclitean Quantum System; as Heraclitus suggested:  the appearance of `things'
depends upon this flux for their continuity and identity.

 We have not discussed quantum `matter': matter will be   excitations
embedded in the spatial quantum structures. Space is not passive, it is not a container.  The
container role of the classical manifolds will finally emerge when we integrate out the HQS
fine detail, resulting in the traditional local field theory formulations.

 We  thank Susan Gunner for her support.

\end{document}